# Meandering gate edges for breakdown voltage enhancement in AlGaN/GaN HEMTs

Sandeep Kumar, Surani B. Dolmanan, Sudhiranjan Tripathy, Rangarajan Muralidharan, Digbijoy N. Nath

*Abstract*— **In this letter, we report on a unique device design strategy for increasing the breakdown voltage and hence Baliga Figure of Merit (BFOM) of III-nitride HEMTs by engineering the gate edge towards the drain. The breakdown of such devices with meandering gate-drain access region (M-HEMT) are found to be 62% more compared to that of conventional HEMT while the ON resistance suffers by 76%, leading to an overall improvement in the BFOM for by 28%. 3D-TCAD simulations show that the decrease in the peak electric field at the gate edge was responsible for increased breakdown voltage.**

## I. Introduction

THE continuous research and development in the field of III-Nitrides in the last three decades had enabled technologies which were difficult to realize with conventional semiconductors [1][2][3][4][5]. One of the promising technologies realized by Nitride semiconductors is high power HEMT switches. The desired properties for a switch are low on-resistance and high breakdown voltage. Highly conducting 2DEG and wide bandgap of III-N (AlGaN/GaN) enable low on-resistance and high breakdown voltages respectively. As the HEMTs are lateral devices, the electric field distribution is not uniform in the off-state condition. The electric field attains maxima at the gate edge towards the drain which leads to breakdown and current collapse. Several field plate designs (gate field plate [6], source field plate [7] and slant field plate [8][9]) were implemented to reduce the peak electric field at the gate edge which had resulted in improved uniformity of electric field distribution and hence increased breakdown voltage. In addition to field plates engineering, Schottky drain contacts were also found to increase the breakdown voltage [10]. Currently, to push the breakdown voltage further, high Al composition AlGaN channels are being investigated [11]. Here, we report a novel approach to increase the breakdown voltage by engineering the gate edge, creating a meandering access region between the drain and the gate.

We acknowledge funding support from MHRD through NIEIN project, from MeitY and DST through NNetRA.
S. Kumar, R. Muralidharan, and Digbijoy N. Nath are with Centre for Nano Science and Engineering (CeNSE), Indian Institute of Science (IISc), Bengaluru, India (e-mail: sandeepku@iisc.ac.in).

## II. Device Fabrication

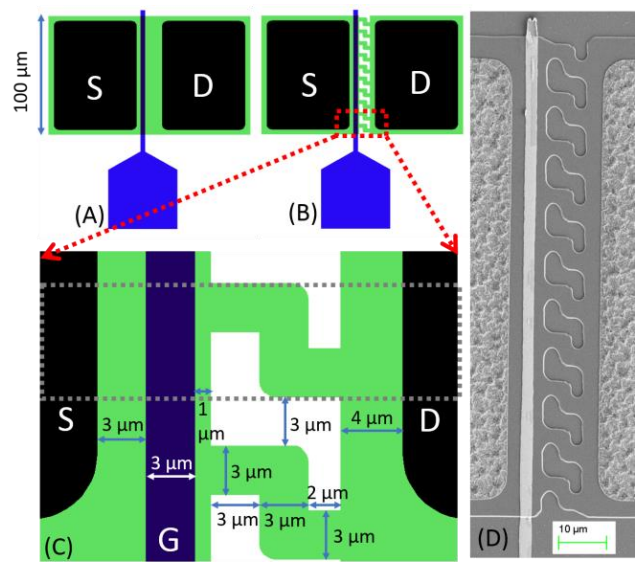

Fig. 1. (A) Schematic diagram of a regular HEMT (B) Gate edge engineered meandered device (M-HEMT) (C) Zoomed image of gate edge region showing various dimensions. AlGaN barrier and the presence of 2DEG is shown as green color. The region in the grey dashed box (one unit) was used for TCAD simulation. Total device width (100 μm) consists of 10 such units. (D) SEM image of device.

The ~5.5 μm thick HEMT stack on 1 mm p-type Si(111) was grown by metal organic chemical vapor deposition (MOCVD)[3]. The unintentional C-doped GaN buffer layer of thickness ~1500 nm was grown prior to overgrowth of the top two-dimension electron gas (2DEG) structures. The top HEMT layers consist of a 300 nm GaN channel layer, a thin ~1.0 nm AlN spacer, ~19 nm $Al_xGa_{1-x}N$ barrier layer with $x = 24\%$, and a ~1 nm thin GaN cap. Hall measurements on the epi-wafer center to edge regions showed an average sheet resistance of 377 ohm/□ with a sheet carrier concentration of about $1.1 \times 10^{13}$ cm$^{-2}$.

Device fabrication started with e-beam evaporation of Ti/Al/Ni/Au metals for Ohmic layer which was then annealed at 850°C for 30 s in $N_2$ ambient. Mesa isolation was done using Cl based reactive ion etching. Device fabrication process ended with e-beam evaporation of Ni/Au metals for gate layer. Two types of devices were fabricated 1) conventional/regular device (C-HEMT) 2) meandered gate edge engineered device (M-HEMT) (Fig. 1) on the same samples side by side. For the fabrication of M-HEMT, the AlGaN barrier and GaN channel were etched out in a particular meandering geometry during mesa etching step as shown in the Fig. 1 (C).

### III. DEVICE CHARACTERIZATION

The devices were characterized for transfer characteristics as shown in the Fig. 2. Both types of devices exhibited the same threshold voltage ($V_{TH}$) as the 2DEG density below the gate remains the same for both devices; however, the reduction in the on-current in the M-HEMT device should be attributed to an increase in the reduction of the effective gate width or periphery. The theoretically estimated value of $V_{TH}$ was found to be -4.64 V (for 21 nm of barrier with barrier dielectric constant of 9) which was much less than the experimentally observed $V_{TH}$ (~-2.5 V). The observed difference in $V_{TH}$ should be attributed to the partial depletion of 2DEG below the Ni/Au Schottky gate [12].

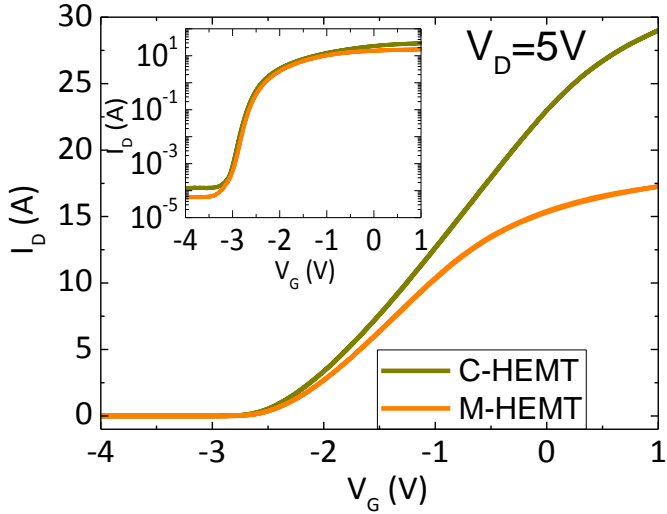

Fig. 2. Transfer characteristics in liner and log scale (inset) of the fabricate devices showing different on-current and same $V_{TH}$. The estimated $I_{ON}/I_{OFF}$ ratio was $10^5$ for both devices.

The output characteristics of the devices are shown in Fig. 3. The static $R_{ON}$ was found to increase by ~1.7x for the M-HEMTs as compared to C-HEMT.

The devices were further characterized for three terminal breakdown (Fig. 4). During the measurements, Silicone oil (s d fine-chem limited) was poured on the sample to avoid contact arcing at high drain biases. The leakage current (gate and drain current) of C-HEMT and M-HEMT were found to reach 10 μA/mm at drain bias of 400 V and 650 V respectively, indicating that M-HEMTs offer a substantially higher breakdown voltage. The leakage currents of both the devices were normalized with 100 μm of device width. The superior breakdown characteristics of M-HEMT were quantitively investigated by the estimation of BFOM which can be defined for 2DEG as: $V_{BD}^2/R_{ON}\left(=qn_s\mu_N E_c^2\right)$ where,

$V_{BD}$ is breakdown voltage, $R_{ON}$ is on-resistance, q is electronic charge, $n_s$ is 2DEG density, $\mu_N$ is electron mobility and $E_C$ is critical electric field. For the estimation of on-resistance of the M-HEMT, the access region includes the etched-out region and 2DEG region. The estimated BFOM of C-HEMT and M-HEMT were found to be $9.3\times10^7$ and $1.3\times10^8$ $V\Omega^{-1}cm^{-2}$ respectively. The BFOM for M-HEMT was found to be higher by 28% in comparison to the C-HEMT. Although these are not record numbers - primarily due to a large gate length - the new design for M-HEMT is expected to push the records when implemented on HEMT samples with lower sheet resistance, Ohmic regrowth and with scaled gate lengths. The superior breakdown characteristics of M-HEMT were further investigated using the 3-dimensional TCAD simulation, which is discussed in the next section.

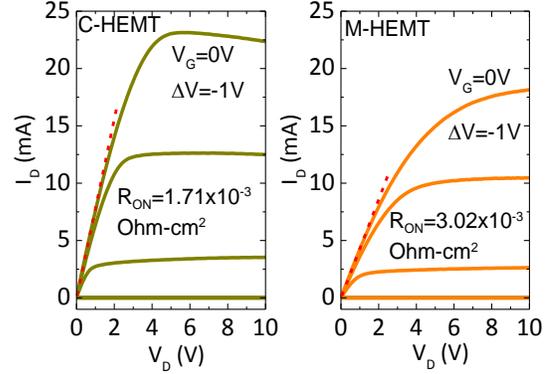

Fig. 3. Output characteristics of devices showing reduction in the on-current and increase in the on-resistance ($R_{ON}$) in M-HEMT device compared to C-HEMT.

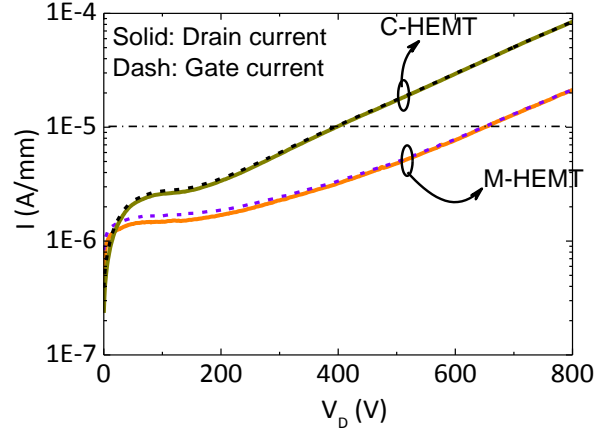

Fig. 4. Three terminal breakdown characteristics of C-HEMT and M-HEMT. The breakdown voltage of M-HEMT was found higher than the C-HEMT by 250 V.

### IV. TCAD SIMULATIONS

Silvaco-ATLAS TCAD tool was used for 3-dimensional device simulation. To reduce the computation load, a unit device width (Fig. 1 (C), shown in the grey dashed box) with a thin HEMT stack (~424 nm) was used for simulation. The material properties were defined by the default parameter lists kp.set2 and pol.set2 of Silvaco ATLAS. The simulated device structure consisted of 24 nm AlGaN barrier, 200 nm GaN channel and 200 nm of GaN buffer. The barrier and channel were unintentionally doped ($1\times10^{15}$ cm$^{-3}$) while a trap density of $1\times10^{15}$ cm$^{-3}$ was considered in the GaN buffer [13]. The simulated electric field profiles of the devices at drain bias of 600 V are shown in the Fig. 5 (A) and (B). The electric field was found to attain a maximum value at the gate edge for both HEMTs, however, the electric field profiles along the device width (ZZ' line) were significantly different for both devices. The Electric field profile along the device width (ZZ' line) for the C-HEMT had not shown any variation while for the M-HEMT, it was found decreasing (Fig. 5 (C)). Also, the peak electric field in the M-HEMT was found lower than that of C-HEMT at the gate edge. The increase in breakdown voltage of M-HEMT in comparison with C-HEMT should be related to the reduction of peak electric field. The simulation overestimates the peak electric field in M-HEMT as in the actual device etched region lies on the both sides of current channel.

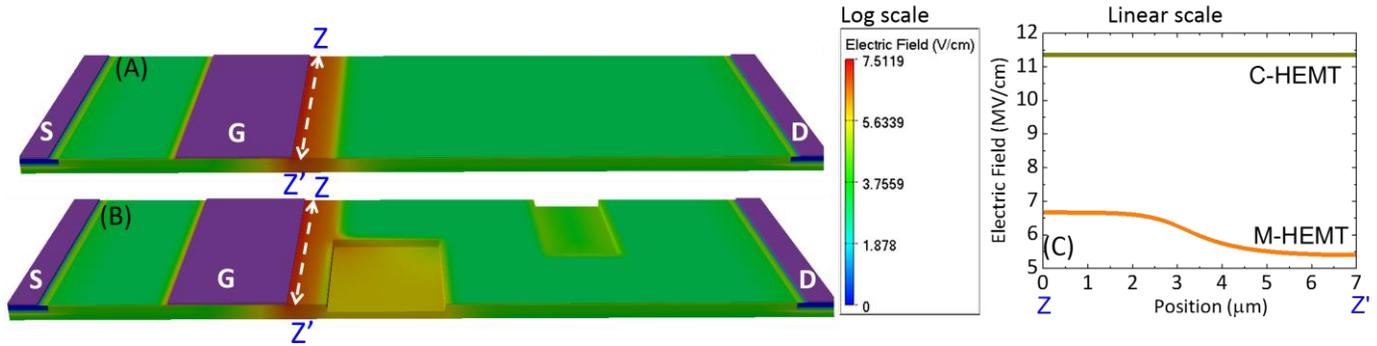

Fig. 5 Electric field profile in the off-state of device at drain bias of 600 V for (A) C-HEMT (B) M-HEMT. (C) the electric field profile along the Z-Z' line for C-HEMT and M-HEMT

## V. Conclusions

A new type of HEMT device was demonstrated which had exhibited superior breakdown characteristics and BFOM compared to a conventional HEMT device. The increase in breakdown voltage should be attributed to decrease in electric field strength at the gate edge due to the unique device structure. As the M-HEMT devices had exhibited promising results further analysis on both the experimental and simulation front are needed to find the optimum device geometries. Also, the application of field plates and/or dielectric passivation needs to be explored to boost the device performance. The simulation presented here was done for ~400 nm thick HEMT stack which is very low compared to the practically used stack due to computation constraints. Hence, a rigorous simulation is also required for thick HEMT stacks.

## VI. Acknowledgement

This publication is an outcome of the Research and Development work undertaken in the Project under Ph.D. scheme of Media Lab Asia. Authors would like to acknowledge the National NanoFabrication Centre (NNFC) and Micro and Nano Characterization Facility (MNCF) at CeNSE, IISc for device fabrication and characterization.